\documentstyle[fleqn,twoside]{article}

\def\noi{\noindent}

\newcommand{\Arthead}[5]{ \setcounter{page}{#4}\thispagestyle{empty}\noi
    \unitlength=1pt \begin{picture}(500,40)
        \put(0,58){\shortstack[l]{\small\it Gravitation \& Cosmology,
                        \small\rm Vol. #1 (#2), No. #3, pp. #4--#5\\
        \footnotesize \copyright \ #2 \ Russian Gravitational Society} }
    \end{picture}
	 }

\newcommand{\Title}[1]{\noi {\Large #1} \\}

\newcommand{\Author}[2]{\noi{\large\bf #1}\\[2ex]\noindent{\it #2}\\}

\newcommand{\Rec}[1]{\noi {\it Received #1} \\}

\newcommand{\Abstract}[1]{\vskip 2mm \begin{center}
        \parbox{16.4cm}{\small\noi #1} \end{center}\medskip}

\newcommand{\foom}[1]{\protect\footnotemark[#1]}

\newcommand{\foox}[2]{\footnotetext[#1]{#2}}

\newcommand{\email}[2]{\footnotetext[#1]{e-mail: #2}}

\newcommand{\Ref}[1]{Ref.\,\cite{#1}}

\def\nq{\hspace*{-1em}}

\def\nqq{\hspace*{-2em}}

\def\cm{\hspace*{1cm}}

\def\inch{\hspace*{1in}}

\def\Jl#1#2{{\it #1\/} {\bf #2},\ }

\def\CQG#1 {\Jl{Clas. Qu. Grav.}{#1}}

\def\DAN#1 {\Jl{Dokl. AN SSSR}{#1}}

\def\GC#1 {\Jl{Grav. \& Cosmol.}{#1}}

\def\GRG#1 {\Jl{Gen. Rel. Grav.}{#1}}

\def\JETF#1 {\Jl{Zh. Eksp. Teor. Fiz.}{#1}}

\def\JMP#1 {\Jl{J. Math. Phys.}{#1}}

\def\NP#1 {\Jl{Nucl. Phys.}{#1}}

\def\PLA#1 {\Jl{Phys. Lett.}{#1A}}

\def\PLB#1 {\Jl{Phys. Lett.}{#1B}}

\def\PRD#1 {\Jl{Phys. Rev.}{D\ #1}}

\def\PRL#1 {\Jl{Phys. Rev. Lett.}{#1}}

\newcommand{\eqsection}{\makeatletter

	\@addtoreset{equation}{section}

	\renewcommand{\theequation}{\arabic{section}.\arabic{equation}}

	\makeatother}

\def\lal{&&\nqq {}}

\def\eqs{Eqs.\,}

\def\beq{\begin{equation}}

\def\eeq{\end{equation}}

\def\bear{\begin{eqnarray}}

\def\bearr{\begin{eqnarray} \lal}

\def\ear{\end{eqnarray}}

\def\earn{\nonumber \end{eqnarray}}

\def\nnn{\nonumber\\ \lal }

\def\tst{\textstyle}

\def\fract#1#2{{\tst\frac{#1}{#2}}}

\def\half{{\fract{1}{2}}}

\def\e{{\,\rm e}}

\def\d{\partial}

\def\const{{\rm const}}

\newcommand{\aver}[1]{\langle \, #1 \, \rangle \mathstrut}

\def\ul{\foox 1 {Talk presented at the Second School-Seminar ``Problems of
Theoretical Cosmology'', Ulyanovsk, Russia, September 2000.}}
\def\Recul{\Rec{31 March 2001}}

\begin{document}

\twocolumn[
\Arthead{7}{2001}{3 (27)}{211}{214}

\Title{ORIGIN OF A CLASSICAL SPACE IN QUANTUM COSMOLOGIES\foom 1}

   \Author{A.A. Kirillov\foom 2 and G.V. Serebryakov\foom 3}
          {Institute for Applied Mathematics and Cybernetics,
		10 Ulyanova St., Nizhny Novgorod 603005, Russia}

\Recul

\Abstract
{The influence of vector fields on the origin of
classical space in quantum cosmologies and on a possible
compactification process in multidimensional gravity is investigated. It is
shown that all general features of the transition between the classical and
quantum evolution regimes can be obtained within the simplest
Bianchi-I model for an arbitrary number of dimensions. It is shown that the
classical space appears when the horizon size reaches the smallest of the
characteristic scales (the characteristic scale of inhomogeneity or a
scale associated with vector fields). In the multidimensional case the
presence of vector fields completely removes the initial stage of the
compactification process which takes place in the case of vacuum models
\cite{k95}. }


]

\ul

\email 2 {kirillov@unn.ac.ru}
\email 3 {gvs@focus.nnov.ru}

\noi
One of the most important problems of quantum cosmology is an adequate
description of the origin of the classical background space. Indeed,
the transition from a pure quantum regime to a quasiclassical one
forms the initial properties of the early Universe and therefore determines
whether the subsequent stages contain an inflationary period and which
kind of initial quantum state should be chosen for its realization
\cite{linde90}. In vacuum inhomogeneous models, the origin of
classical background space was first investigated in \Ref{KM97}.
It was found that the background metric belongs to the class of
quasi-isotropic spaces, and the origin of the classical space
corresponds to the instant when the horizon size matches the
inhomogeneity scale of the space. However, a complete
investigation of the problem requires studying the influence of
different matter sources and also reserving the possibility that
our Universe has extra dimensions, as predicted by a number of unified
theories \cite{string}. In the latter case, the transition between
quantum and classical regimes should include the so-called
compactification stage \cite{comp}. And indeed, as was shown in
Refs.\,\cite{k95}, quantum evolution of inhomogeneous models in
dimensions smaller or equal than ten includes an initial stage of such a
compactification. In this case the initial expansion of the Universe
proceeds in an anisotropic way when, along the extra dimensions, scales
decrease, and the expansion only occurs in three dimensions.

In this paper we study the influence of vector fields on the origin of
classical space and on the possible compactification process in
multidimensional gravity. It turns out that, in general, classical
space appears when the horizon size reaches the minimal one among the
scales:  the characteristic inhomogeneity scale or the characteristic
scale associated with the vector fields, which is an analogue of Jeans'
wavelength (in the present paper we do not consider the inhomogeneous
case, and therefore we shall discuss the second possibility only).
However, in multidimensional case the presence of vector fields
completely removes the initial compactification stage, the process
found in Refs.\,\cite{k95}. This may be a good reason why
the vector fields should not be included as external fields, but should
rather be composed from additional metric components in the
compactification process.

First, we note that the origin of a background space is not a
specific\ problem for quantum cosmology only. Such a problem does also
exist in classical theory when the gravitational field and matter
sources are described by a probabilistic measure distribution with
unstable statistical properties. Just this case was shown to be
realized in approaching the cosmological singularity \cite{K93}.
It turns out that in both cases (classical or quantum cosmology) the
background space formation mechanism is the same, while the
nature of the statistical description is different.  Moreover, estimates for
the formation time of the background differ by a factor which
collects all quantum corrections and depends on the choice of the initial
conditions (and also on the specific scheme of quantization). This ensures
the correctness of our consideration and provides a hope that whatever
quantum gravity will be constructed in the future, our results will
survive, (save, probably, minor corrections).

Let $A_{\mu }=( \varphi, A_{\alpha }) $ be a vector field
($\alpha =1,2,\ldots ,n$), and for the metric we shall use the standard
decomposition
\beq
	ds^{2}=N^{2}dt^{2}-g_{\alpha \beta }( dx^{\alpha }+N^{\alpha }dt)
	( dx^{\beta }+N^{\beta }dt) . 	\label{met}
\eeq
Then the action takes the form (in what follows we use the Planckian units)
\bearr
	I=\int d^{n}x dt \biggl\{ \pi^{\alpha \beta }\frac{\d }{\d t}
	g_{\alpha \beta }+\pi ^{\alpha }\frac{\d }{\d t}A_{\alpha}
\nnn \cm
	+\varphi \d _{\alpha }\pi ^{\alpha }-NH^{0}-N^{\alpha }H_{\alpha}
		\biggl\} , \label{act}
\ear
where
\bearr \nq
 H^0=\frac{1}{\sqrt{g}}\left\{ \pi _\beta^\alpha \pi _\alpha^\beta
 	-\frac{1}{n-1}( \pi _\alpha^\alpha) ^2 + \frac{1}{2}
   g_{\alpha \beta }\pi ^{\alpha }\pi ^{\beta }+V\right\} , \label{hamcn}
\\ \lal
 H_{\alpha}=-\nabla _{\beta }\pi _{\alpha }^{\beta }+\pi^{\beta}
 		F_{\alpha\beta }, \label{momcn}
\ear
here $F_{\alpha \beta }=\d _{\alpha }A_{\beta }-\d _{\beta} A_{\alpha }$, \
$V=g( \frac{1}{4}F_{\alpha \beta }F^{\alpha \beta }-R)$, and $R$
is the scalar curvature with the metric $g_{\alpha \beta }$. Varying the
action with respect to $\varphi $, we find the constraint $ \d _{\alpha }\pi
^{\alpha }=0$, so it suffices to consider only the transverse parts for
$A_{\alpha }$ and $\pi ^{\alpha }$. Thus in what follows we set $\varphi =0$
and $\d _{\alpha }\pi ^{\alpha }=0$ to be satisfied.

It is convenient to use the so-called generalized Kasner-like
parametrization of the dynamical variables \cite{K93,k95}. The metric
components and their conjugate momenta are represented as follows:
\beq
	 g_{\alpha \beta }=\sum_{a}\exp \left\{ q^{a}\right\}
	 \ell_{\alpha}^{a}\ell _{\beta}^{a}\,,\qquad \pi _{\beta }^{\alpha
	}=\sum_{a}p_{a}L_{a}^{\alpha }\ell _{\beta }^{a}\,, \label{ksnpr1}
\eeq
where $L_{a}^{\alpha }\ell _{\alpha }^{b}=\delta _{a}^{b}$ ($a,b=0,...,(n-1)$
), and the vectors $\ell _{\alpha }^{a}$ contain only $n(n-1)$ arbitrary
functions of the spatial coordinates. A further parametrization may be taken
in the form
\beq
	\ell _{\alpha }^{a}=U_{b}^{a}S_{\alpha }^{b},\qquad U_{b}^{a}\in
	SO(n),\qquad S_{\alpha }^{a}=\delta _{\alpha }^{a}+R_{\alpha }^{a}
	\label{ksnpr3}
\eeq
where $R_{\alpha }^{a}$ denotes the  triangle matrix ($R_{\alpha }^{a}=0$ as
$ a<\alpha $ ). Substituting \eqs (\ref{ksnpr1}), (\ref{ksnpr3}) into
(\ref{act} ), we find the following expression for the action functional:
\bearr
	I=\int_{S}(p_{a}{\frac{\d q^{a}}{\d t}}+T_{a}^{\alpha }
	{\frac{\d R_{\alpha }^{a}}{\d t}}+\pi ^{\alpha }\frac{\d
	A_{\alpha }}{\d t}
\nnn \cm
	-NH^{0}-N_{\alpha }H^{\alpha })d^{n}x\,dt,    	\label{actksnr}
\ear
where $T_{a}^{\alpha }=2\sum_{b}p_{b}L_{b}^{\alpha }U_{a}^{b}$, and the
Hamiltonian constraint acquires the structure
\bearr
  H^{0}={\frac{1}{\sqrt{g}}}\biggl\{ \sum p_{a}^{2}-{\frac{1}{n-1}}
  	\biggl(\sum p_a\biggr)^{2}
\nnn  \inch
	+\frac{1}{2}\sum \e^{q^a}( \pi ^{a}) ^{2}+V \biggr\} .\label{hcnstr}
\ear
In the last equation the potential $V$ collects all spatial derivatives, and
we have used the notation $\pi^a = \sum \pi^{\alpha }\ell _{\alpha }^{a}$.
In case $n=3$, the functions $R_{\alpha }^{a}$ are only connected by
transformations of a coordinate system and may be removed by resolving the
momentum constraints $H^{\alpha }=0$ \cite{K93}. However, in the
multidimensional case the functions $R_{\alpha }^{a}$ contain
$\half n(n-3)$ dynamical functions as well.

It can be shown that near the singularity, in the case $n>3$, all spatial
derivatives can be neglected in the leading order. Therefore, we neglect the
potential $V$ (the terms $F_{\alpha \beta }$ and $R$) in the action
(\ref{hcnstr}). In the case $n=3$, the curvature term cannot be neglected.
However, in this case the kinetic term of the vector field
($\frac{1}{2}g_{\alpha \beta }\pi ^{\alpha }\pi ^{\beta }$) induces
precisely the same type of evolution (from qualitative and even
quantitative viewpoints) as that of the curvature terms.

Thus, in this approximation, the Einstein equations formally coincide with
the equations for homogeneous Bianchi-I model, and it is a remarkable fact
that, near the singularity, a homogeneous Bianchi-I model with a vector
field contains all qualitative features of the general inhomogeneous models.
In what follows we shall use the gauge $N^{\alpha }=0$ and, for the sake of
simplicity, consider a homogeneous model, i.e., all functions depend
on time only. We also use the normalization of the space volume $V^{n}=
\int_{S}d^{n}x=1$. Thus we find (within our approximation) the equations for
the vector field
\bearr
	E_{\alpha }=\frac{\d }{\d t}A_{\alpha }=\frac{N}{\sqrt{g}}
	g_{\alpha \beta }\pi ^{\alpha },
\\ \lal
	\frac{\d }{\d t}\pi ^{\alpha }=0.
\ear
The last equation gives $\pi ^{\alpha }=\const$, and in what follows we shall
treat the quantities $\pi ^{\alpha }$ as external parameters (which are
actually eigenstates of the respective operators). The rest of the present
paper repeats mainly the method suggested in \Ref{KM97}. Near the
singularity, it is convenient to make use of the following parametrization of
the scale functions \cite{K93}:
\beq
	q^{a}=\ln R^{2}+Q_{a}\ln g;\cm \sum Q_{a}=1,\, \label{SF}
\eeq
where we distinguish a slow function of time $R$, which characterizes the
absolute value of the metric functions \cite{M69,Grav} and is specified by
initial conditions (see below), and the anisotropy parameters $Q_{q}$ and $
\ln g=\sum q^{a}-2n\ln R$ can be expressed in terms of the new set of
variables $\tau ,$ $y^{i}$ ($i=1,2...(n-1)$), as follows:
\beq\nq
	 Q_{a}\left( y\right) =\frac{1}{n}\left(
	 1+\frac{2y^{i}A_{i}^{a}}{1+y^{2}} \right) ,\quad
	 \ln g=-n\e^{-\tau }\frac{1+y^{2}}{1-y^{2}} 	\label{AP}
\eeq
where $A_{i}^{a}$ is a constant matrix, see, e.g., Refs.\,\cite{K93}. The
parametrization (\ref{AP}) has the range $y^{2}<1$ and $-\infty <\tau
<\infty $ ($0\leq g\leq 1$), and an appropriate choice of the function $R$
allows one to cover by this parametrization the whole classically allowed
region of the configuration space.

The evolution (rotation) of the Kasner vectors results in a slow time
dependence of the functions $\pi ^{a}$, and it can be shown that these
functions are completely determined by the momentum constraints, while the
evolution of the scale functions is described by the action
\bearr
	I=\int \biggl\{\biggl(\vec{P}}\frac{\d \vec{y}}{\d t}
			+h{\frac{\d \tau}{\d t}\biggr)
\nnn  \ \
        -\frac{N}{n( n-1) R^{n}\sqrt{g}}\e^{2\tau}
	[\varepsilon ^{2}-h^{2}+U(\tau ,\vec{y})]\biggr\}dt, \label{Act}
\ear
where ${\varepsilon }^{2}=\frac{1}{4}( 1-y^{2})^2 {\vec{P}}^{2}$,
and the potential term $U$ (which comes from the kinetic term of the vector
field) has the following structure:
\beq
	U=n(n-1) R^{2}\e^{-2\tau }\sum_{a=1}^{n}\left( \pi ^{a}\right)
		^{2}g^{Q_{a}}. \label{V}
\eeq
Here the coefficients $\pi ^{a}$ (projections of $\pi ^{\alpha }$ on
the Kasner vectors) are slow functions of $\ln g$ and characterize the
initial intensity of the vector field. In the approximation of deep
oscillations, when $g\ll 1$ , this potential can be modelled by a set of
potential walls:
\beq
	g^{Q_{a}}\rightarrow \theta _{\infty }[Q_{a}]=\left\{
	\begin{array}{ll}
		+\infty \,\,,\,\, & Q_{a}<0, \\
		0\,\,\,,\qquad & Q_{a}>0,
	\end{array}
		\right. \label{Apr}
\eeq
and is independent of the Kasner vectors
$U_{\infty }=\sum \theta _{\infty}( Q_{a}) $.

By solving the Hamiltonian constraint $H=0$ in (\ref{Act}) we determine the
ADM action \cite{ADM}, reduced to the physical sector as follows:
\beq
    I=\int ({\vec{P}}_{\vec{y}}\cdot \frac{d\vec{y}}{d\tau }-H_{ADM})d\tau ,
		\label{5b}
\eeq
where $H_{ADM}\equiv -h=\sqrt{\varepsilon ^{2}+U}$ is the ADM Hamiltonian
and $\tau $ plays the role of time ($\dot{\tau}=1$), which corresponds to the
gauge
\[
        N_{ADM}=n( n-1) R^n\sqrt{g}[2H_{ADM}]^{-1}\e^{-2\tau }.
\]

The applicability condition of the approximation (\ref{Apr}) can be
written as follows:
\beq
	\varepsilon ^{2}\gg U \label{in}
\eeq
as $Q_{a}>\delta >0$ ($\delta \ll 1$). Thus, from the condition that the
approximation of deep oscillations (\ref{Apr}) breaks at the instant
$g\sim 1$, one finds that the function $R$ should be chosen as follows:
$R^{2}=\varepsilon ^2 [n( n-1) \pi ^2]^{-1} \e^{2\tau }$
(where $\pi ^{2}=\sum ( \pi ^{a})^2$), and the inequality (\ref{in}) reads
$ g\ll 1$.

The synchronous cosmological time is related to $\tau $ by means of the
equation $dt=N_{ADM}d\tau $, from which we find the estimate $\sqrt{g}\sim
t/t_{0}$, where $t_{0}=cL^{n}\varepsilon ^{n-1}$, (cf. \Ref{KM97}), $L\sim
1/\pi $ is a characteristic scale related to the vector field, $\varepsilon
$ is the ADM energy density ($\varepsilon =\const$), and $ c$ is a slow
(logarithmic) function of time ($c\sim 1$ as $g\to 1$ ). Thus, in
the synchronous time, the upper limit of the approximation (\ref {Apr}) is
$t\sim t_{0}$. We note that from the physical viewpoint $t_0$ corresponds
to the instant when the horizon size reaches the characteristic scale related
to the energy of the vector field, and both terms in the Hamiltonian
constraint (the kinetic energies of the anisotropy and of the vector field)
acquire the same order.

The physical sector of the configuration space (the variables $\vec{y}$) is a
realization of the Lobachevsky space, and the potential $U_{\infty }$
bounds the part $K$ $=\{ Q_{a}\geq 0\} $. Quantization of this
system can be carried out as follows. The ADM energy density represents a
constant of motion, and therefore we can define stationary states as
solutions to the eigenvalue problem for the Laplace--Beltrami operator $
-\varepsilon ^{2}=\Delta +{\frac{( n-2)^2}{4} P}$ (see Refs. \cite{K97,k95})
\bearr
        \biggl(\Delta +k_{J}^{2}+{\frac{\left( n-2\right) ^{2}}{4}P}\biggr)
        \varphi_J(y)=0,\qquad \varphi _{J}\Big|_{\d K}=0, \label{eq:3.7}
\nnn
\ear
where the Laplace operator $\Delta $ is constructed via the metric $\delta
l^{2}=h_{ij}\delta y^{i}\delta y^{j}= 4(\delta  y)^2 (1-y^2)^{-2}$.
The eigenstates $\varphi _{J}$ are classified by the integer number $J$ and
obey the orthogonality and normalization relations
\beq
        (\varphi _{J},\varphi _{J'})=\int_{K}\varphi _{J}^{\ast }(y)
                \varphi _{J'}(y)D\mu (y)=\delta _{JJ'}, \label{eq:3.8}
\eeq
where $D\mu (y)=\frac{1}{a_{n}}\sqrt{h}d^{2}y\left( x\right) $ and $a_{n}$
is the volume of $K$. Thus, an arbitrary solution $\Psi $ to the
Schr\"odinger equation $i\d _{\tau }\Psi =H_{ADM}\Psi $ takes the form

\beq
        \Psi =\sum_{J}\exp (-ik_{J}\tau )\varphi _{J}(y)C_{J} \label{eq:3.11}
\eeq
where $C_{J}$ are arbitrary constants which are to be specified by initial
conditions. Within our approximation, these constants also represent
arbitrary functions of the vector field
\beq
        C_{J}( A) =\int d^{n-1}\pi C_{J,\pi }\exp (i\pi^{\alpha}A_{\alpha}),
\eeq
and the normalization condition reads ($\int d^{n-1}A\sum_{J}\mid
C_{J}( A) \mid ^{2}=1$). The probabilistic distribution of the variables $y$
has the standard form $P( y,\tau ) =| \Psi ( y,\tau ) | ^2$. The eigenstates
$\varphi _{J}$ determine stationary (in terms of the anisotropy parameters
$Q( y) $) quantum states and describe an expanding universe with a fixed
energy of the anisotropy.

For an arbitrary quantum state $\Psi $ we can determine the background
metric $\aver{ds^2}$. However, such a background is stable and has sense
only when quantum fluctuations around it are small. In case $g\ll 1$,
fluctuations well exceed the average metric, and the background is hidden
\cite{K97}. Indeed, consider the moments of scale functions
$\aver {a_i^M} =\aver{ R^M g^{\frac{M}{2}Q_i}}$. The leading contribution
in $\aver{ a_i^M}$ comes from those points of the billiard at which $ Q_{i}$
takes a minimal value. Such points are at the boundary of the billiard, and
the minimal value of $Q$\ is $Q_{i}^{\min }=0$. Since $\varphi_{J}(\d K)=0$,
in the neighbourhood of $\d K $ we have $\varphi _{J}\approx \eta _{J}Q$,
and the probability density is
\bearr
	P_{\tau }( Q) =\int_{K}P( y,\tau ) \delta (Q-Q( y))
		\sqrt{h}d^{n-1}y
\nnn \cm
                \approx B_{J}( \tau) Q^{f( n) }
\ear
as $Q\to 0$. Here $f( n) =2$ for $n>3$ and $f(3) =3/2$ (since in case
$n=3$ we get $\sqrt{h}\sim 1/\sqrt{Q}$).  Thus, in the limit $g\to 0$,
the moments of the function $\aver{a_i^M}$ are given by ($M>0$)
\beq
   \aver{ a_i^M} \simeq D_{i}( M,\tau ) \frac{1}{( M\ln 1/g_{\ast })
   	 ^{f( n) +1}}, \label{moments}
\eeq
where $g_{\ast }=g( \tau ,y^\ast ) $ and $D_{i}$ is a function slowly
varying in time, which collects the information of the initial quantum state.

Consider now an arbitrary stationary state $\varphi _{J}$ which gives the
stationary probabilistic distribution $P( y) =| \varphi_{J}| ^{2}$.
In this case $D\simeq $ $bk_{J}^{M}( L^{M})_{J}$
is a constant, where $b$ comes from the uncertainty in the operator
ordering. Thus, for the intensity of quantum fluctuations one finds the
divergent, in the limit $g\to 0$ ($\tau \to -\infty $), expression
$\aver{ \delta ^2} = ( \aver{a^2} /\aver{a}^2 - 1) \sim (\ln 1/g_{\ast })
^{f( n)+1}$, which explicitly shows the instability of the average geometry
as $g\ll 1$. The intensity of quantum fluctuations reaches the order $\delta
\sim 1$ at the instant $t\sim ( L^{n})_{J}k_{J}^{n-1}$ ($g\sim 1$)
when the anisotropy functions can be described by small perturbations
$a_{i}^{2}=R^{2}( 1+Q_{i}\ln g+\ldots ) $, and the Universe acquires a
quasi-isotropic character.  This instant can be considered as
the origin of a stable classical background. It is important that strong
anisotropy of a stable classical geometry turns out to be forbidden. We
recall that the hypothesis that the very beginning of the Universe evolution
should be described by quasi-isotropic models, while anisotropic models are
forbidden, was first suggested in \Ref{LS74} from a different consideration
(as a result of the impossibility of constructing a self-consistent theory
which could account for back reaction of particle creation in such models).

To conclude, we make two important remarks. Firstly, in the presence of the
vector field, the evolution of the metric undergoes spontaneous
stochastization for an arbitrary number of dimensions \cite{dh}. This follows
from the fact that potential walls always restrict a finite region of the
configuration space (Vol.$( K) <\infty $ for arbitrary $n$), and
the metric should be described by an invariant measure. In this case,
estimates for the moments of scale functions can be obtained by setting
$P( y,\tau ) =1$, which gives the replacement $f(n)\to f(n)-2$ in
(\ref{moments}). Thus the above picture of the formation of a stable
background remains to be valid in classical theory as well (with minor
corrections of the estimates). And secondly, the presence of a vector
field drastically changes the structure of the configuration space in
dimensions $n>3$. The billiard turns out to be finite for an arbitrary
number of dimensions \cite {dh}, but the payment is the fact that the
initial stage of the compactification process, which was found in vacuum
models \cite{k95}, is absent. Indeed, in the model considered above, the
anisotropy parameters have the range $1\geq Q\geq 0$ and remain always
positive (we recall that in the vacuum case these functions could have
negative values, e.g., they changed in the range $1\geq Q\geq
-(n-3)/(n+1)$). This means that the expansion always proceeds in such a way
that the lengths monotonicaly increase in all spatial directions. This
probably represents a rather serious reason why vector fields should not
be included as external fields in multidimensional gravity.


This research was supported by RFBR (Grant No. 98-02-16273) and DFG 
(Grant No. 436 Rus 113/23618).


\end{document}